\documentclass[a4paper,fleqn,usenatbib]{mnras}

\usepackage{newtxtext,newtxmath}
\usepackage[T1]{fontenc}
\usepackage{ae,aecompl}

\usepackage{graphicx}	
\usepackage{amsmath}	
\usepackage{amssymb}	
\usepackage{siunitx}

\title[$FUV$ Observation of the GC NGC 6397]{Far-Ultraviolet Observation of the Globular Cluster NGC 6397}

\author[A. Dieball et al.]{
A. Dieball$^{1,2}$\thanks{E-mail: adieball@astro.uni-bonn.de},
A. Rasekh$^{3}$,
C. Knigge$^{4}$,
M. Shara$^{5}$,
D. Zurek$^{5}$
\\
$^{1}$Argelander Institut f{\"u}r Astronomie der Universit{\"a}t Bonn, Auf dem H{\"u}gel 71, 53121 Bonn, Germany \\
$^{2}$Helmholtz-Institut f{\"u}r Strahlen und Kernphysik, Nussallee 14-16, 53115 Bonn, Germany \\
$^{3}$Oskar Klein Centre, Department of Astronomy, Stockholm University,SE-106 91 Stockholm, Sweden\\
$^{4}$Physics and Astronomy, University of Southampton, Southampton SO17 1BJ, UK\\
$^{5}$Department of Astrophysics, American Museum of Natural History, New York, NY 10024, USA\\
}

\date{Accepted XXX. Received YYY; in original form ZZZ}

\pubyear{2016}

\begin{document}
\label{firstpage}
\pagerange{\pageref{firstpage}--\pageref{lastpage}}
\maketitle

\begin{abstract}
  We present an observational far-UV ($FUV$) and near-UV ($NUV$) study
  of the core region of the globular cluster NGC\,6397. The
  observations were obtained with the Space Telescope Imaging
  Spectrograph (STIS, FUV), and the Wide Field Camera 3 (WFC3, NUV) on
  board the Hubble Space Telescope. Here, we focus on the $UV$ bright
  stellar populations such as blue stragglers (BSs), white dwarfs
  (WDs) and cataclysmic variables (CVs). We present the first
  $FUV-NUV$ color-magnitude diagram (CMD) for this cluster. To support
  our classification of the stellar populations, we compare our
  $FUV-NUV$ CMD with optical data from the ACS Survey of Galactic
  Globular Clusters. The $FUV-NUV$ CMD indicates 16 sources located in
  the WD area, and ten BSs within the $25''\times 25''$ of the STIS
  $FUV$ data. Eighteen Chandra X-ray sources are located within the
  $FUV$ field of view. Thirteen of those have a $NUV$ counterpart, of
  which nine sources also have a $FUV$ counterpart. Out of those, five
  sources are previously suggested CVs, and indeed all five are
  located in the WD/CV region in our $FUV-NUV$ CMD. Another CV only
  has a $FUV$ but no $NUV$ counterpart. We also detect a $NUV$ (but no
  $FUV$) counterpart to the MSP located in the core of this
  cluster. The $NUV$ lightcurves of the CVs and MSP show flickering
  behaviour typical of CVs. We found that the BSs and CVs are the most
  centrally concentrated population. This might be an effect of mass
  segregation or indicate the preferred birth place of BSs and CVs via
  dynamical interactions in the dense core region of GCs. HB stars are
  the least centrally concentrated population and absent in the
  innermost area of the core.
\end{abstract}


\section{Introduction}

Globular clusters (GCs) are the oldest and most massive stellar
aggregates in our Galaxy. The stellar density in the cores of GCs can
go up to $10^6$ stars pc$^{-3}$. 
Therefore, close encounters or even direct collisions with resulting
mergers between the cluster members can be quite frequent, resulting
in dynamically formed new, exotic stellar populations such as blue
stragglers (BSs), cataclysmic variables (CVs), low-mass X-ray binaries
(LMXBs), millisecond pulsars (MSPs) and other close binary (CB)
systems.  CBs are important for the evolution of GCs, since the
binding energy of a few CBs can be comparable to the binding energy of
the entire GC. By transferring their orbital energy to passing single
stars, CBs can have a significant impact on the dynamical evolution of
the cluster \citep[see][and references therein]{Elson1987,Hut1992}.
\cite{Heggie2009} and \cite{Giersz2009} performed N-body and Monte
Carlo simulations of the two closest GCs NGC\,6397 and M\,4 to study
the evolution of their central structure. Both clusters are similar in
mass, but show different surface brightness profiles. The simulations
suggest that both clusters are core-collapsed, but the different
surface brightness profiles can be explained by fluctuations of the
core, which are influenced by dynamically active binaries
\citep{Heggie2009}. Interacting CBs like CVs are an important subset
of CBs and can be used to trace the overall CB population in a
cluster. CVs may form via ordinary evolution of primordial binaries,
and this formation mechanism can already account for significant
numbers of CVs \citep[see, e.g.,][]{Davies1997}, but dynamical
formation scenarios like tidal capture and three body encounters can
also create CVs in the dense cores of GCs. For example,
\cite{DiStefano1994}, \cite{Shara2006}, and \cite{Ivanova2006}
discussed several primordial and dynamical formation scenarios and
predicted a few 100 CVs in a GC like 47\,Tuc. \cite{Pooley2006}
suggested that the CVs in dense GCs are predominantly dynamically
formed.

How can we study the CV and other exotic stellar populations in a GC
observationally? Most exotica like CVs and LMXBs are optically faint,
and the cores of GCs, their birth sites, are crowded, making their
detection even more difficult. However, such exotic stellar
populations show a spectral energy distribution that peaks at short
wavelengths, often in the near-ultraviolet ($NUV$) or far-ultraviolet
($FUV$). This makes the $FUV$ the ideal waveband to search for and
study exotic sources such as BSs, CVs, blue and extreme horizontal
branch (HB) stars, and LMXBs. In addition, main sequence (MS) stars
and red giants, which make up the majority of the stars in a GC, are
faint at such short wavelengths. As a result, $FUV$ images tend to be
considerably less crowded than optical images.  So far, deep {\it HST}
$FUV$ observations to study the exotic stellar populations in GCs have
been conducted in only a few clusters, namely 47 Tuc
\citep{Knigge2002,Knigge2008}, NGC 2808
\citep{Brown2001,Dieball2005,Servillat2008}, M15
\citep{Dieball2005M15,Dieball2007}, NGC 1851
\citep{Zurek2009,Zurek2016,Maccarone2010}, M80
\citep{Dieball2010,Thomson2010} and NGC 6752 \citep{Thomson2012}.

Interacting CBs are also strong X-ray emitters, and indeed large
numbers of X-ray binaries have been detected in a number of GCs with
the {\it Chandra} X-ray telescope \citep[e.g.][]{Grindlay2001a,
  Grindlay2001b, Pooley2002, Edmonds2003a, Edmonds2003b,
  Heinke2003809H, Heinke2003516H, Hannikainen2005, Lugger2007,
  Elsner2008, Bogdanov2010, Cohn2010, Forestell2014}. The combination
of both X-ray and $FUV$ observations is thus an ideal approach to
detect and study the interacting CBs in GCs.

NGC 6397 is the second closest GC to us, at 2.2 kpc \citep{Heyl2012},
and is one of the best studied GCs. \cite{Hansen2013} suggested a
metallicity of [Fe/H]=-1.8 dex and reddening of E$_{B-V}$ = 0.18
mag. \cite{Husser2016} provided 3D spectroscopy of more than 12,000
stars in NGC\,6397 and derived a mean metallicity of [Fe/H]=-2.12 dex,
compared to [Fe/H]=-2.02 dex listed in the catalogue of Harris (1996,
updated version 2010). Absolute proper motions and the cluster's orbit
around the Galaxy have been measured by \cite{Kalirai2007} and
\cite{Milone2006}. \cite{Lind2011} investigated the chemical
composition of RGB stars in NGC\,6397 based on high-resolution VLT
FLAMES spectra and suggested that the cluster contains a primordial
($\approx$ 30\%) and a secondary (70 \%) generation that formed from
material enriched by the first generation. They further suggested that
the current mass of the cluster is just 10\% of its initial
mass. Multiple generations were photometrically confirmed by
\cite{Milone2012}, who found that the MS in NGC 6397 splits into two
components, suggesting the presence of two stellar
generations. \cite{Heyl2012} studied the cluster kinematics and
estimate a cluster mass of $1.1 \times 10^5 M_\odot$.

\cite{Richer2008} derived a deep optical CMD that nearly reaches the
H-burning limit and showed, for the first time, the blue-turn and the
truncation of the WD sequence in this cluster at $F814W \approx 28$
mag. The mass function (MF) agrees with a flat power-law or log-normal
distribution, which in turn agrees with the predictions from N-body
modelling presented in \cite{Hurley2008}. The latter conclude that, in
different fields in the cluster, the WD population did not change
strongly due to dynamical processes, whereas the MF of the MS has been
altered significantly by the cluster dynamics.

Comparing the observed CMD and WD cooling sequence with theoretical
models, \cite{Torres2015} suggested an age of 12.8 Gyr, whereas
\cite{Hansen2007} derived a younger (also WD based) age of 11.47
Gyr. \cite{Torres2015} used updated evolutionary tracks and suggested
that this might be the reason for the difference in age compared to the
findings from \cite{Hansen2007}. These values bracket the MS turnoff
age of 12 Gyr that has been suggested by both \cite{Gratton2003} and
\cite{Winget2009}.  Comparing the CMDs of NGC\,6397 with 47\,Tuc and
M\,4, \cite{Richer2013} found that NGC\,6397's MS is much bluer (as
expected because of its lower metallicity), but the location of the WD
sequences agrees independent of metallicity. This is expected as WD
atmosphere should be composed of only H and He.
Thus, the location of the (bottom of the) WD sequence provides an
ideal age indicator.

\cite{Cool1995} first searched for CVs in NGC\,6397 using $R$- and
$H\alpha$-band {\it HST} WFPC imaging. They found three H-alpha
emission objects, all of them are also UV bright and ROSAT X-ray
counterparts, making them good CV candidates. {\it Chandra} X-ray
observations have been carried out by \cite{Grindlay2001a} and
\cite{Bogdanov2010}, detecting 79 X-ray sources within the half-mass
radius of this cluster. \cite{Shara2005} used WFPC2 and STIS UV
imaging and detected dwarf nova (DN) eruptions in two CVs that had
previously been suggested to be magnetic systems \citep{Grindlay2001b}
and which are usually not expected to show DN
eruptions. \cite{Shara2005} suggested that these two systems might be
intermediate polars that occasionally undergo DN eruptions, which
might explain the scarcity of DNs in GCs.

More recently, \cite{Cohn2010} carried out much deeper optical and
$H\alpha$ {\it HST} imaging, finding 69 optical counterparts to the 79
X-ray sources. Of these, 15 are CVs or CV candidates and two are
MSPs. Optical emission of the bright and centrally concentrated CVs
seems to be dominated by their accretion disks and secondaries,
whereas the fainter CVs are optically dominated by their WD primary.
A sequence of 24 He WDs, parallel to the CO WD cooling sequence, has
also been suggested to be present in the cluster's central region by
\cite{Strickler2009}.

Recently, \cite{Kamann2016} used integral field spectroscopic data and
suggested that the surface brightness profile and velocity dispersion
profile could either be explained by the presence of an
intermediate-mass black hole with a mass of 600 $M_{\odot}$ - which
agrees with an earlier suggestion by \cite{deRijcke2006} of $\approx
1300$ to $400 M_{\odot}$, based on radio continuum emission - or with a
dark stellar component.

Here, we present the results of a STIS $FUV$ imaging analysis of the GC
NGC 6397. We compare our data to WFC3 $NUV$ and ACS optical data. The
observations and data reduction are described in Sect.~\ref{obs}, and
the analysis is presented in Sect.~\ref{data}, including the UV and
optical CMDs and a discussion of the various stellar populations. In
Sect.~\ref{nuvvar} we present the $NUV$ lightcurves of the CVs and the
MSP in our $FUV$ field of view. The radial distribution of the various
stellar populations that show up in our UV and optical CMDs are
discussed in Sec.~\ref{radial}. Our results are summarized in
Sect.~\ref{sum}.

\section{Observations and Data Reduction:}
\label{obs}

\subsection{The STIS $FUV$  Data}\label{fuv_catalog}

The $FUV$ data was obtained with the Space Telescope Imaging
Spectrograph (STIS) on-board the {\it HST} using the $FUV$-MAMA
detector and the F25QTZ filter (program GO-8630, PI: M. Shara). The
STIS FUV-MAMA instrument has a field of view of $25''\times 25''$ and
a plate scale of $0.024''$/pixel, and the central wavelength of the
F25QTZ filter is at 1590~\AA. Four F25QTZ images are available with
exposure times of 1200, 2333, 865 and 600 seconds, resulting in a
total exposure time of 4998 seconds.

As a first step we created a $FUV$ master image using the four
pipeline-produced flat-fielded (i.e. calibrated) images. The
flat-fielded images were aligned using the {\tt Tweakreg} task from
the {\tt Drizzlepac} package running under {\tt PyRaf} which computes
the residual shifts between all the input images and the reference
image (the first of our input images). Next, the {\tt AstroDrizzle}
task from the {\tt Drizzlepac} package was used to combine all the
aligned images and produce a geometrically corrected, deep master
image.

\subsubsection{Source Detection}

In order to detect all potential sources, we used the {\tt Daofind}
task in the {\tt Daophot} package \citep{Stetson1987} running under
{\tt PyRAF} to create an initial list of potential sources. We note
that $FUV$ images do not suffer from severe crowding, because MS stars
and red giants, which make the majority of the stars in GCs, are not
expected to be bright at such short wavelengths.

The coordinates of the initial source list were overplotted on the
master image and carefully inspected by eye. Overall, the source
detection routine {\tt Daofind} worked well, but some of the detected
sources appeared to be false detections around bright sources and PSF
spikes and were consequently discarded. On the other hand, some faint
sources could be detected by eye but had not been found by the
detection routine and thus were added by hand to the final list, which
comprises 192 sources.

\subsubsection{Aperture Photometry}

Aperture photometry was performed on the combined and
geometrically-corrected $FUV$ master image using {\tt Daophot}
\citep{Stetson1987} running under {\tt PyRAF}. Since only about 200
sources are present in the $FUV$ master, i.e. the image is not very
crowded, we set an aperture size of 8 pixels, with a small sky annulus
of 8-12 pixels, so that for most sources no neighbouring source is
within the sky annulus. We also allowed for a Gaussian re-centering of
the input coordinates derived from {\tt Daofind}.

However, an aperture radius of 8 pixels does not include 100\% of the
source flux, while on the other hand the small sky annulus means that
not only background, but also source flux will be subtracted during
sky subtraction. Thus, we need to correct for these two
effects. Following \cite{Dieball2007}, we picked a few isolated stars
and measured their curve of growth (flux versus aperture radius) out
to an aperture radius of 57 pixels, as well as their magnitude ratio
using a small (8 - 12 pixel) and a large (60 to 70 pixel) sky
radius. The flux correction factors turned out to be very small, with
an aperture correction $\rm{\tt ApC} = 1.11 \pm 0.06$ and a negligible
sky correction ($\rm{\tt SkyC} = 1.00 \pm 0.00$). The magnitude of
each source was then determined in the STMAG system as

\begin{equation}\label{mag_formula}
  \rm{\tt STMAG} = -2.5 \times \log_{10} ({\tt CR} \times \rm{\tt Photflam} \times \rm{\tt ApC} \times \rm{\tt SkyC}) + {\tt ZPT} 
\end{equation}

where $\rm{\tt Photflam}$ is the inverse sensitivity of the filter
taken from the image headers as $1.19004155 \times 10^{-16}$
ergs/seconds/cm$^2$/\AA\ per counts/seconds, $\rm{\tt CR}$ denotes the
count rate of source, and $\rm{\tt ZPT} = 21.1$ mag is the zeropoint
for the \rm{\tt STMAG} system (also taken from the image header).

\subsection{The $NUV$ data}

The $NUV$ observations were obtained with the Wide Field Camera 3
(WFC3) UVIS/F225W camera/filter combination as part of program
GO-11633 (PI: R. M. Rich). The WFC3/UVIS instrument has a field of
view of $162''\times 162''$ and a plate scale of $0.04''$/pixel. The
data consist of 24 \mbox{images} with an exposure time of 680 seconds
each, resulting in a total exposure time 16320 seconds.

\begin{figure*}
  \centering
  \includegraphics[width=\textwidth]{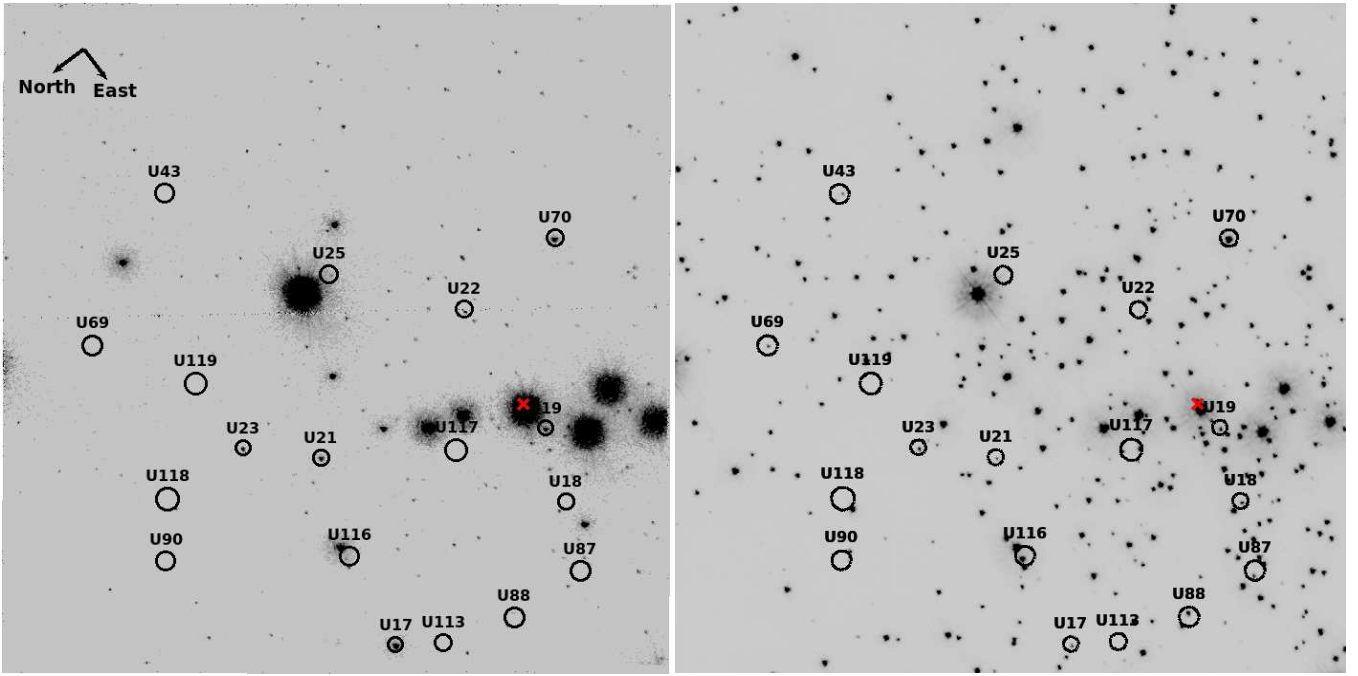}
  \caption{Left: The $FUV$ master image of NGC 6397. Right: A subset
    of the $NUV$ master image covering the same field. Note that the
    full $NUV$ master image covers a much larger field of view
    ($162''\times 162''$), whereas the area covered by the $FUV$ data
    is much smaller ($25''\times 25''$). Both images are displayed on
    a logarithmic scale to bring out the faint sources. The locations
    of the X-ray sources are marked with their 95\% error circles and
    their X-ray denotation from Bogdanov et al. (2010) and Cohn et
    al. (2010). The centre of the cluster, taken from Cohn et
    al.~(2010), is marked with a red cross and is close to the
    location of the X-ray source U\,19 and nearly coincides with the
    location of a bright BS source.}
  \label{FUV_NUV_master_images}
\end{figure*}

The $NUV$ master image was created in the same way as the $FUV$ master
image, using the pipeline produced, flat-fielded images as input to
{\tt Tweakreg} and {\tt AstroDrizzle} and default parameter settings.
The $FUV$ and $NUV$ master images are shown in Figure
\ref{FUV_NUV_master_images}.

\subsubsection{Creating the $NUV$ Catalog}
\label{Creating_the_NUV_Catalog}

As can be seen, the $NUV$ master image is considerably more crowded
than the $FUV$ master image. We therefor use the photometry package
{\tt Dolphot} which is a stellar photometry package adapted from {\tt
  HSTphot} \citep{Dolphin2000} to search for sources and perform
PSF-fitting photometry. In contrast to our $FUV$ aperture photometry,
the source finding and photometry routines are not performed on the
deep $NUV$ master image, but on the individual flat-fielded images,
with the $NUV$ master serving as a reference frame for the image
coordinates.  Photometry is performed on each individual flat-fielded
image using the WFC3 module within {\tt Dolphot}, Jay Anderson's PSF
library for the WFC3/UVIS F225W filter and the corresponding pixel
area map\footnote{which are available on the {\tt Dolphot} web-page:
  americano.dolphinsim.com/dolphot/}. The WFC3 module also includes
built-in charge transfer efficiency (CTE) corrections and a
photometric calibration to the VEGAmag system.  {\tt Dolphot} provides
routines to mask all pixels that are flagged as bad in the data
quality image, to multiply by the pixel area maps, and to align all
the input images to the reference image using user-defined coordinate
lists for each input and the reference image. For all these steps, we
began with the recommended parameters and then optimized them such
that the photometry was pushed as deep as possible. However, the three
brightest stars located in the $FUV$ field of view are saturated in
the $NUV$ data. The {\tt Dolphot} output provides a magnitude from the
individual images, and we used these to determine a mean magnitude for
these three $NUV$ sources, however only for the sake of plotting them
in the $FUV-NUV$ CMD, see Fig.~\ref{cmdfuv}. We caution the reader
that these magnitudes are likely not the correct $NUV$ magnitude, as
{\tt Dolphot} returns an error flag for each individual measurement
(too many saturated pixels). The three stars in question are marked in
Fig.~\ref{cmdfuv}.

\subsection{Optical Catalog}

The main objective of this paper are the STIS $FUV$ data and the
$FUV-NUV$ CMD (see Sect.~\ref{fuvcmd}). However, in order to find
optical counterparts to our $FUV$ sources, we also used some archival
optical data obtained with the Advanced Camera for Surveys (ACS) in
the F606W (corresponding to $V$-band) and F814W (corresponding to
$I$-band) filters (program GO-10775,PI: A. Sarajedini) as part of the
``ACS Survey of Galactic Globular Clusters''
\citep{Sarajedini2007}. Data for each filter consist of one 1 second
exposure and four 15 second exposures, i.e. a total exposure time of
61 seconds. The ACS/WFC instrument has a field of view of $202''\times
202''$ and a plate scale of $0.05''$ per pixel. We used the photometry
catalogue provided via the ``ACS Survey of Galactic Globular
Clusters'' web-page. The data and data reduction are described in
detail in \cite{Sarajedini2007} and \cite{Anderson2008}.

\subsection{The X-ray Catalog}
\label{x_ray}

In order to find matches to known X-ray sources in our $FUV$ field, we
used the catalog published in \cite{Cohn2010}, who in turn used the
X-ray data published in \cite{Bogdanov2010} and
\cite{Grindlay2006}. \cite{Cohn2010} searched for optical counterparts
to the 79 Chandra X-ray sources within the halfmass radius of NGC
6397, and identified 15 CVs, 42 ABs, two MSPs, one active galactic
nucleus, and one interacting galaxy pair candidates. In total, they
found 69 optical counterparts to the 79 X-ray sources.

\subsection{Matching Catalogs}
\label{matching}

All {\it HST} data used in this paper have been obtained at different
times and with slightly different pointings. This has an impact on the
world (RA, DEC) coordinates obtained from the images, i.e. even though
the same source is measured, the derived coordinates typically differ
slightly. For this reason, all coordinates first have to be
transformed to a common coordinate system. We picked the $NUV$ master
image (F225W) as our overall reference image towards which all $FUV$
and optical (pixel) coordinates were transformed. The $NUV$ master
image was chosen because it is not as crowded as the optical data, and
therefore it was easiest to find stars in the $NUV$ master that could
also clearly be identified in the optical and $FUV$ data. We used 108
stars that were easily identified in the $NUV$ and $FUV$ master images
to calculate the transformation coefficients using {\tt Geomap}
running under {\tt PyRAF}. Next, the coordinate transformation was
applied to the whole FUV coordinate list using the {\tt Geoxytran}
task running under {\tt PyRAF} to transform them to the $NUV$
reference coordinate system. For the optical coordinate
transformation, we used 136 stars that can be clearly identified in
the $NUV$ master and the ACS optical catalogue. Again, we first used
{\tt Geomap} to work out the coordinate transformation, and applied
{\tt Geoxytran} to the whole optical coordinate list to transform the
ACS coordinates to our $NUV$ reference system.

In order to find possible matches to the known X-ray sources, we also
transformed the X-ray coordinates taken from Table~1 in
\cite{Cohn2010} to the $NUV$ master coordinate system. Since
\cite{Cohn2010} tied their astrometry to the USNO UCAC2 catalogue, we
used 30 of the 36 USNO UCAC2 stars that are located within the
$NUV$ master field of view and for which a $NUV$ counterpart could be
clearly identified. Again, we use {\tt Geomap} and {\tt Geoxytran} to
work out the coordinate transformation from UCAC2 to our $NUV$
reference frame. We then applied this transformation to the X-ray
source coordinates listed in Table~1 in \cite{Cohn2010}\footnote{Note
  that we first transformed all coordinates from the world coordinate
  system to the $NUV$ master image physical system, i.e. pixel
  coordinates, and all transformations are applied to the
  corresponding pixel coordinates so that all coordinates which are
  used for matching are in the $NUV$ image (physical) reference
  frame.}.

After transforming all coordinates into a common ($NUV$) coordinate
system, we chose a maximum tolerance radius of 2 $NUV$ pixels to
search for counterparts in the $FUV$ and optical. 

We also searched for $FUV$ and $NUV$ counterparts to X-ray sources
within the 95\% positional uncertainty radius given in
\cite{Bogdanov2010}, see Fig.~\ref{FUV_NUV_master_images}. All
possible $FUV$ and $NUV$ counterparts are listed in
Table~\ref{NUVXray}, see Sect.~\ref{cvs} below.

\subsection{False Matches}
\label{false}

The source density in the $NUV$ and optical images is high and these
images are much more crowded than the $FUV$ image. Thus we might
expect a few false matches among our $FUV-NUV$ and $NUV$-optical
pairs. The number of these spurious pairs depends on the number of
sources in the $FUV$ (192 sources), $NUV$ (416 sources) and optical
(678 sources) in the $FUV$ field of view, our chosen matching
tolerance radius (2 $NUV$ pixel), and the number of sources actually
matched (182 $FUV$/$NUV$ pairs, 175 $FUV$/optical and 407
$NUV$/optical pairs). Following the approach in \cite{Knigge2002}, we
expect fewer than one false match (0.14) among the $FUV-NUV$ pairs,
and also fewer than one false match (0.16) among the NUV-optical
pairs.

\section{Data Analysis:}
\label{data}

In this section, we discuss the CMDs and the various stellar
populations that show up, including WDs, CVs, HB stars, BSs and X-ray
counterparts.

\subsection{The $FUV-NUV$ CMD}
\label{fuvcmd}

\begin{figure}
  \centering
  \includegraphics[width=0.5\textwidth]{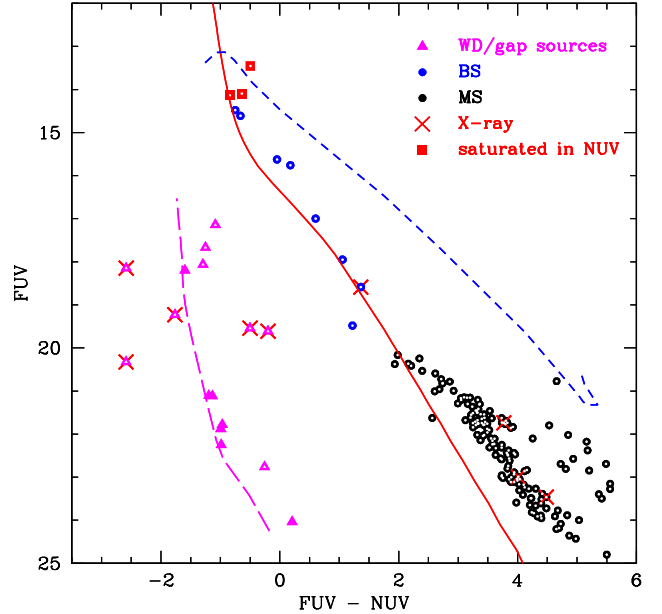}
  \caption{$FUV$ vs. $FUV-NUV$ CMD. For orientation purpose, we also
    show the synthetic ZAHB (blue, short-dashed line), ZAMS (red,
    solid line) and WD cooling sequences (magenta, long-dashed
    line). Sources with optical counterparts are marked with open
    symbols. Large black and blue dots/circles denote MS and BS stars,
    respectively. The three brightest sources are saturated in the
    $NUV$ and are marked with red squares, their true $NUV$ magnitudes
    are likely brighter and hence their colour might also be
    redder. Sources that are possible X-ray counterparts are marked
    with red crosses, and WD/Gap sources, which include CVs, are
    marked with magenta triangles.}
  \label{cmdfuv}
\end{figure}

The $FUV-NUV$ CMD is plotted in Fig.~\ref{cmdfuv}. For orientation
purpose, we also show the location of a theoretical zero-age MS (ZAMS,
red line), a zero-age horizontal branch (ZAHB, blue short-dashed
line), and a WD cooling sequence (magenta long-dashed line). In order
to create a ZAMS, we used the fitting formulae provided by
\cite{Tout1996} to estimate appropriate stellar parameters. We then
interpolated on the grid of Kurucz models and folded the resulting
synthetic spectra through the response of the appropriate
filter+detector combinations using {\tt Synphot} running under {\tt
  PyRAF}. For the ZAHB track, the grid of oxygen-enhanced ZAHB models
provided by \cite{Dorman1992} was used to generate a set of models
closest to the cluster metallicity. Again, the corresponding $FUV$ and
$NUV$ magnitudes were computed by interpolating on the Kurucz model
grid and folding through the corresponding filter transmission curves
using {\tt Synphot}. We used standard cluster parameters of $E_{B-V} =
0.18$ mag, a distance of 2.2 kpc, and a metallicity of
[Fe/H]=-2.02~dex \citep{Harris1996}, which gives a good fit of the
theoretical sequences to the underlying data.

We can see a relatively tight sequence of sources (small black data
points/circles) reaching from the red, faint corner at $FUV\approx 24$
mag towards bluer and brighter magnitudes, with only eleven sources
(blue circles and red squares) brighter than 20 mag in the $FUV$ along
the ZAMS. The faint, red sequence consists mostly of MS stars, whereas
the eleven bright sources agree with being BSs or possibly blue HB
stars. The three brightest sources are saturated in the $NUV$ and we
assign a mean magnitude based on the magnitudes in the individual
input images {\it only for plotting purpose}. Their true $NUV$
magnitude is likely brighter and as a consequence they might be
redder. At first glance, it is not clear whether these three sources
are bright BSs or possibly blue HB stars.

Sixteen sources are located on the blue side of our ZAMS. This is
where we expect WDs and gap sources \citep[i.e. sources that are
located in the gap between the ZAMS and the WD cooling sequence,
see][]{Knigge2002}. For comparison, we computed a theoretical WD
sequence by interpolating on the \cite{Wood1995} grid of theoretical
WD cooling curves, adopting a mean WD mass of $0.55 M_{\odot}$. The
models were translated to the observational plane by carrying out
synthetic photometry with {\tt PySynphot}, using a grid of synthetic
DA WD spectra provided by \cite{Gaensicke1995} with temperatures
ranging from 60,000~K to 10,000~K. It seems that the hottest WDs in our
$FUV-NUV$ CMD have temperatures around 48,000~K, corresponding to
$FUV\approx17$ mag, whereas the coolest WD (candidates) are around
$FUV\approx 24$ mag, corresponding to $T_{eff} \geq 10,000$~K.

\subsection{Optical CMD}

As can be seen in Fig.~\ref{cmdfuv}, the ZAHB and ZAMS cross at
$FUV\approx13$ mag. Based on just the $FUV-NUV$ CMD, it is not
immediately clear whether the three brightest sources are BSs or
$FUV$-bright blue HB stars. In order to address this issue, we
searched for optical counterparts to all $FUV$ sources which helps
to distinguish between the various stellar populations. 

\begin{figure}
  \centering
  \includegraphics[width=0.5\textwidth]{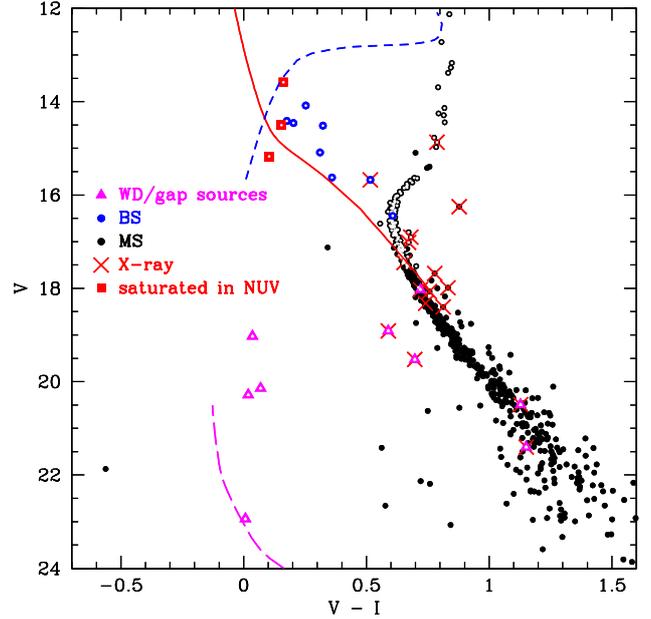}
  \caption{Optical CMD of NGC 6397, data taken from the ACS Survey of
    Galactic Globular Clusters (Sarajedini et al.~2007). The symbols
    are the same as in Fig.~\ref{cmdfuv}, i.e. WDs are marked with
    magenta triangles, BSs with blue circles (selected from the
    $FUV-NUV$ CMD in Fig.~\ref{cmdfuv}), the three sources saturated
    in the NUV are marked with red squares, and possible counterparts
    to X-ray sources are denoted with red crosses. All optical sources
    that have a $FUV$ counterparts are marked with open symbols. We
    also overplot the same theoretical sequences as in
    Fig.~\ref{cmdfuv}. Note that this CMD covers the FUV field of view
    only, not the full ACS field of view.}
  \label{cmdopt}
\end{figure}

\begin{figure}
  \centering
  \includegraphics[width=0.5\textwidth]{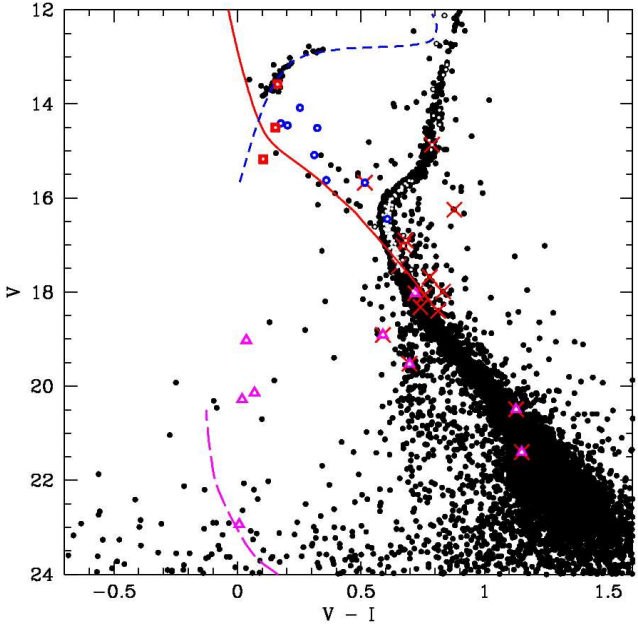}
  \caption{The same as Fig.~\ref{cmdopt}, but for the full optical
    field of view. The symbols and theoretical tracks are the same as
    in Fig.~\ref{cmdfuv} and ~\ref{cmdopt}.}
  \label{cmdoptfull}
\end{figure}

The optical CMD is based on the ACS survey \citep{Sarajedini2007} and
plotted in Fig.~\ref{cmdopt}. Again, BSs are marked in blue, the three
sources saturated in the $NUV$ are marked with red squares, and WD/gap
sources are denoted with magenta triangles. Counterparts to X-ray
sources are marked with red crosses. All $FUV$ sources that have an
optical counterpart, and vice versa all optical sources that have a
$FUV$ counterpart, are indicated with open symbols in
Fig.~\ref{cmdfuv} and Fig.~\ref{cmdopt}. We also overplot the
theoretical sequences discussed in Sect.~\ref{fuvcmd}, which fit the
underlying CMD well. However, again it is not conclusive whether the
three sources marked with red squares are the brightest BSs in the
field of view, or blue HB stars. Note that we only show sources that
are located within the $FUV$ field of view. Thus, for comparison, we
show the optical CMD from the full ACS field of view in
Fig.~\ref{cmdoptfull}. Again, we marked those optical sources with a
$FUV/NUV$ counterpart with open symbols. Based on this CMD, it appears
that the brightest of the three sources saturated in the $NUV$ is
indeed a blue HB star. The other two sources are most likely bright
and blue BSs, but could also be fainter (and slightly redder) extreme
blue HB (EHB) stars. A classification of these three sources was not
possible based on the $FUV-NUV$ CMD alone.

\begin{figure}
  \centering
  \includegraphics[width=0.5\textwidth]{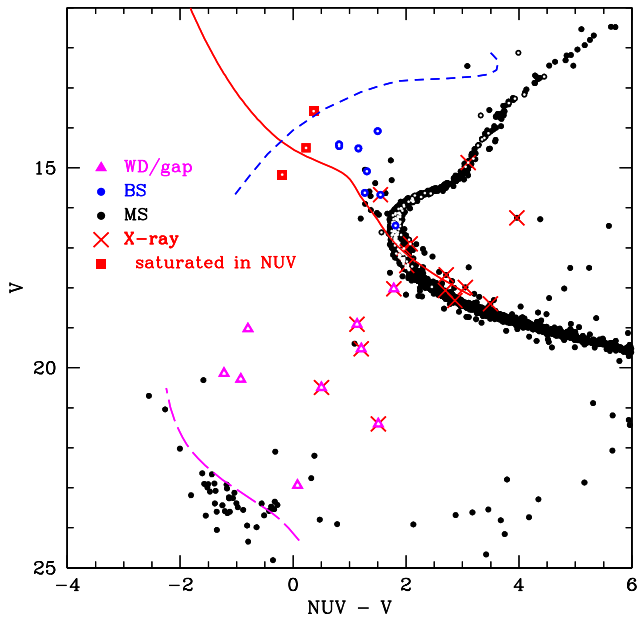}
  \caption{$NUV -V$ CMD, note that this CMD covers a larger area than
    the $FUV$ master image, but the $NUV$ and optical images do not
    fully overlap. The symbols and theoretical tracks are the same as
    in Figs.~\ref{cmdfuv}, ~\ref{cmdopt} and \ref{cmdoptfull}. A WD
    sequence can be clearly seen.}
  \label{cmdnuvopt}
\end{figure}

\subsection{Horizontal Branch Stars}
\label{hb_analysis}
 
Only one, possibly up to three HB stars seem to be located within the
$FUV$ field of view which covers only the inner $25''\times25''$ of
the cluster, see Fig.~\ref{cmdfuv} and \ref{cmdopt}. Thus, HB stars
seem to be rare in the very core region of this cluster. This is
likely an effect of the dynamical evolution of this cluster and mass
segregation, i.e. more massive stars travel towards the cluster
centre, but less massive stars, like the HB stars which typically have
masses of $\approx 0.5 M_{\odot}$, are pushed out of the dense cluster
centre \citep[see e.g.][who derived an average mass of $\approx 0.6
M_{\odot}$ for the HB population in M\,80]{Dieball2010}.

44 HB stars are located within the full ACS field of view, which
covers $202\arcsec \times 202\arcsec$ (and indeed somewhat more,
because four images were taken per filter during the ACS observations,
shifted by $\approx4\arcsec$ in x and y direction). Assuming an even
distribution of the HB stars across the field (i.e. ignoring the
concentration of stars towards the cluster centre as well as effects
of mass segregation), and scaling this number to the field of view
covered by the STIS $FUV$ observation, we would expect less than one
HB star.  

Next, we try to take the central concentration into account by
assuming that HB stars have a distribution more similar to the MS
stars. We find 3136 MS stars within the full ACS field of view, out of
those, 588 (18.7\%) are located within a $19.8\arcsec$ (corresponding
to 500 NUV pixel) radius around the cluster centre, 83 stars (2.6\%)
within a $5.5\arcsec$ (130 NUV pixel) radius, and 345 (11\%) within
the $FUV$ field of view (these numbers are based on the ACS optical
CMD, down to $V = 20$ mag, see Fig.~\ref{cmdoptfull}, and see also
Sect.~\ref{radial}). Assuming that the HB stars follow a distribution
similar to the MS stars, this translates to $8^{+12.7}_{-5.9}$ HB
stars within the 500 pixel radius (we found six HB stars),
$1^{+7.9}_{-1}$ HB star within the 130 pixel radius (we found none),
and about $5^{+11}_{-4.2}$ HB stars within the full FUV field of view
(we found one HB star), where the errors are the $3\sigma$ confidence
limits taken from Gehrels (1986, Tables 1 and 3). The probability of
finding one or less HB stars within the FUV field of view is 3.2\%
according to Bernouille statistics. This is not significant.

Thus, the small number of HB stars found in our
$FUV$ field of view could as well simply be a statistical effect.

\subsection{Blue Stragglers}
\label{bs_analysis}

Blue stragglers are peculiar stellar systems, since they are not
following standard single star evolution. They are found in all GCs in
the Galaxy, and are located above the MS turnoff in the CMDs but close
to the ZAMS, i.e. they occupy a region in the CMD that should be
devoid of stars. Their CMD location indicates that they are more
massive than MS stars, and they also seem to be more centrally
concentrated \citep[e.g.][]{Ferraro2003}, another indication of their
higher than average mass. The formation mechanism of BSs is still a
matter of debate, but there are two generic scenarios for these
systems: BSs might form from the collision of two MS stars resulting
in a rejuvenated and more massive (MS) star \citep{Hills1976}, or they
might have formed via the evolution of primordial binaries
\citep{McCrea1964}.

A well-defined trail of eleven stars above the MS turn-off along (but
slightly redder than) the ZAMS can be seen in Fig.~\ref{cmdfuv} and
\ref{cmdopt}, marked in blue (and red for the three sources saturated
in the $NUV$). This is the expected location of the BS population. As
discussed above in Sec.~\ref{hb_analysis}, one to three of the
brightest of these stars might be blue HB stars.

We can identify 29 BSs in the full optical CMD, however, only ten of
those are located inside the field of view of our $FUV$ data (assuming
that only the brightest $FUV$ source is a HB star). Nine of our $FUV$
selected BSs are optically bright, the remaining $FUV$ bright BS
is optically faint and located in the MS turn-off region. Its $FUV$
excess might be explained with a WD companion which contributes flux
to the $FUV$ \citep[see][who found BS-WD binaries in the old open
cluster NGC\,188]{Gosnell2015}

\subsection{White Dwarfs/Gap Sources}

A population of 16 sources close to the theoretical WD cooling
sequence can be seen in our $FUV-NUV$ CMD in Fig.~\ref{cmdfuv}. The
scatter of sources around the bright, hot part of the WD sequence is
large, so we cannot distinguish single WDs, He WDs or binary systems
based on the FUV-NUV CMD alone. However, CVs are interacting binaries
containing a WD accreting from a low-mass companion star. They are
expected to be located between the WD cooling sequence and the ZAMS. 

The $NUV-V$ CMD is displayed in Fig.~\ref{cmdnuvopt}. Again, we show
the theoretical sequences, and we can clearly see a trail of blue and
faint data points that agree nicely with the theoretical WD
sequence. Sources that have a $FUV$ counterpart are marked with open
symbols. Only nine WD/gap sources have a $FUV, NUV$ and optical
counterpart. In the $NUV-V$ CMD, they are all located between the WD
cooling sequence and the MS, whereas in the $FUV-NUV$ CMD, three of
these sources are located on the blue side of the $FUV$ WD cooling
sequence. These three sources are also X-ray sources, suggesting that
they are interacting binaries, i.e. CVs, and indeed their location in
the $NUV-V$ CMD supports this classification.

\subsection{X-ray counterparts}
\label{cvs}

CVs are also X-ray sources, so we searched for counterparts to known
X-ray sources in the inner area of NGC\,6397, listed in
\cite{Cohn2010}. In total, there are 18 X-ray sources in the $FUV$
field of view. The 95\% positional uncertainty radius of these X-ray
sources were taken from \cite{Bogdanov2010}. The position of the X-ray
sources are marked in Fig.~\ref{FUV_NUV_master_images} and we searched
for counterparts within their 95\% confidence radii. 

Nine out of the 18 X-ray sources had both a $FUV$ and a corresponding
$NUV$ counterpart, namely U17, U19, U21, U22, U23, U70, U88, U116 and
U117.  Five of these, the counterparts to U17, U19, U21, U22, and U23,
are located in the WD or gap area in the $FUV-NUV$ CMD, which makes
all of them good CV candidates. Indeed, \cite{Cohn2010} also suggested
that these sources are CVs. In their $H\alpha-R$ CMD (their Fig.~3),
these five sources are bright $H\alpha$-excess sources, i.e. they are
optically bright CVs. Their optical counterparts in Fig.~\ref{cmdopt}
are located on or slightly blue-ward of the MS and are all fainter than
$V < 18$ mag.

The counterpart to U70 is a bright $FUV/NUV$ source and is located on
the ZAMS in the BS area above the MS turnoff in our $FUV-NUV$
CMD. This source was also identified as a BS in \cite{Cohn2010}.
This is certainly an interesting and peculiar object
that deserves future observation.

The remaining possible counterparts to U88, U116 and U117 are all
located on the red side of the MS in all CMDs, i.e. they are likely AB
stars. This agrees with the classification of \citet[][their Table 1
and Fig. 4]{Cohn2010}. The $UV$ counterpart to U88 is located just at
the rim of its error circle. For U116 and U117, two $NUV$ sources are
located within their error circles. For U116, only the brighter of the
two $NUV$ sources has a $FUV$ counterpart which places this source on
the MS in the $FUV-NUV$ CMD, marked with a red cross. Both $NUV$
sources have an optical counterpart (marked with red crosses in the
optical and NUV-optical CMDs), however, one is located on the RGB and
the other on the optical MS. \cite{Cohn2010} identified a MS star as
the counterpart to U116, thus, the brighter source on the RGB might
not be the real counterpart but the fainter source on the MS (which
has, however, no $FUV$ match). For U117, both of the two $NUV$ sources
are close to the rim of the X-ray error circle. Only one has a $FUV$
counterpart, but both have an optical counterpart which are both
marked with a cross in the optical CMD in Fig.~\ref{cmdopt}. They are
located close to each other on the MS, but we cannot decide which
might be the real counterpart based on the CMDs alone.

X-ray source U25 only has a $FUV$ counterpart but no $NUV$ source was
detected. However, there seems to be a faint smudge at the position in
the $NUV$ master image, but no magnitude could be assigned. Since this
source does not appear in our $FUV-NUV$ CMD, we cannot attempt any
further classification, however \cite{Cohn2010} classified U25 as a
possible CV. 

U18 was classified as a likely MSP by \cite{Cohn2010} and
\cite{Bogdanov2010}. We can clearly detect a $NUV$ counterpart,
however, the $FUV$ master image only shows a faint smudge at the
position of the $NUV$ source. 

Also sources U43, U69, and U87 only had a $NUV$ counterpart, but no
$FUV$ counterparts could be detected. Thus, we cannot classify these
sources, however \cite{Cohn2010} classified all three sources as
ABs. For source U43, two $NUV$ sources and corresponding optical
counterparts are located within its X-ray error circle. Both sources
are located on the MS in the optical CMD. The brighter $NUV$ source is
closer to the centre of the error circle and thus might be the true
counterpart.

No $UV$-counterparts have been found for the X-ray sources U90, U113,
U118 and U119. The $NUV$ master image shows a faint smudge within the
U90 error circle, but no measurement was returned by our Dolphot
photometry. U90 is classified as an AB, and U118 is classified as a
MSTO star in \cite{Cohn2010}, but sources U113 and U119 still need to
be classified.

The X-ray sources with a $FUV$ and/or $NUV$ counterpart that are
located inside the $FUV$ field of view are listed in
Table~\ref{NUVXray}.

\begin{table*}
\begin{center}
  \caption{\label{NUVXray} $FUV$ and/or $NUV$ counterparts to the
    X-ray sources that are located within the $FUV$ field of view. The
    $V$ and $I$ magnitudes are taken from the optical catalogue
    provided by the ACS survey of Galactic Globular Clusters, see
    Sarajedini et al.~(2007). The last column gives our classification
    as a CV, BS or MS source, based on the position of the counterpart
    in the $FUV-NUV$ and/or optical CMD. The $^1$ denotes the
    classification derived in this paper, a $^*$ denotes the
    classification given in Cohn et al.~(2010), which agrees well with
    ours. See the text for details.}
\begin{tabular}{cccccccccc}
\hline
Id$_{Xray}$ & $FUV$ & $\Delta FUV$ & $NUV$ & $\Delta NUV$ & $FUV-NUV$ & $V$ & $V-I$ & $NUV-V$ & type\\
\hline
 U17  & 18.147 & 0.016 & 20.732 & 0.005 &-2.585 & 19.525 & 0.697 & 1.207 & CV$^{1,*}$\\
 U18  &   -    &   -   & 20.214 & 0.004 &   -   & 16.250 & 0.878 & 3.964 & MSP$^*$\\
 U19  & 19.542 & 0.038 & 20.043 & 0.004 &-0.501 & 18.911 & 0.589 & 1.132 & CV$^{1,*}$\\
 U21  & 19.231 & 0.026 & 20.996 & 0.006 &-1.765 & 20.493 & 1.128 & 0.503 & CV$^{1,*}$\\
 U22  & 20.326 & 0.043 & 22.913 & 0.017 &-2.587 & 21.403 & 1.151 & 1.510 & CV$^{1,*}$\\
 U23  & 19.612 & 0.031 & 19.810 & 0.003 &-0.198 & 18.026 & 0.720 & 1.784 & CV$^{1,*}$\\
 U25  & 21.475 & 0.128 &   -    &   -   &   -   &   -    &   -   &   -   & CV?$^*$\\
 U43  &   -    &   -   & 21.040 & 0.006 &   -   & 17.989 & 0.832 & 3.051 & MS$^1$,AB$^*$\\
 U43  &   -    &   -   & 21.186 & 0.006 &   -   & 18.319 & 0.741 & 2.867 & MS$^1$(?)\\
 U69  &   -    &   -   & 20.766 & 0.005 &   -   & 18.070 & 0.757 & 2.696 & AB$^*$\\
 U70  & 18.587 & 0.019 & 17.224 & 0.001 & 1.363 & 15.676 & 0.516 & 1.548 & BS$^1$,AB/BS$^*$\\
 U87  &   -    &   -   & 20.389 & 0.004 &   -   & 17.679 & 0.778 & 2.710 & AB$^*$\\
 U88  & 23.464 & 0.182 & 18.980 & 0.002 & 4.484 & 16.904 & 0.681 & 2.076 & MS$^1$,AB$^*$\\
 U116 & 21.741 & 0.120 & 17.972 & 0.001 & 3.769 & 14.871 & 0.788 & 3.101 & MS$^1$(?)\\
 U116 &   -    &   -   & 21.894 & 0.010 &   -   & 18.400 & 0.813 & 3.494 & AB$^*$\\
 U117 &   -    &   -   & 19.460 & 0.003 &   -   & 17.447 & 0.653 & 2.013 & AB?$^*$\\
 U117 & 22.992 & 0.147 & 18.981 & 0.002 & 4.011 & 17.017 & 0.672 & 1.964 & MS$^1$,AB$^*$(?)\\
\hline
\end{tabular}
\end{center}
\end{table*}

\section{$NUV$ Variability}
\label{nuvvar}

\begin{figure*}
  \centering
  \includegraphics[width=\textwidth]{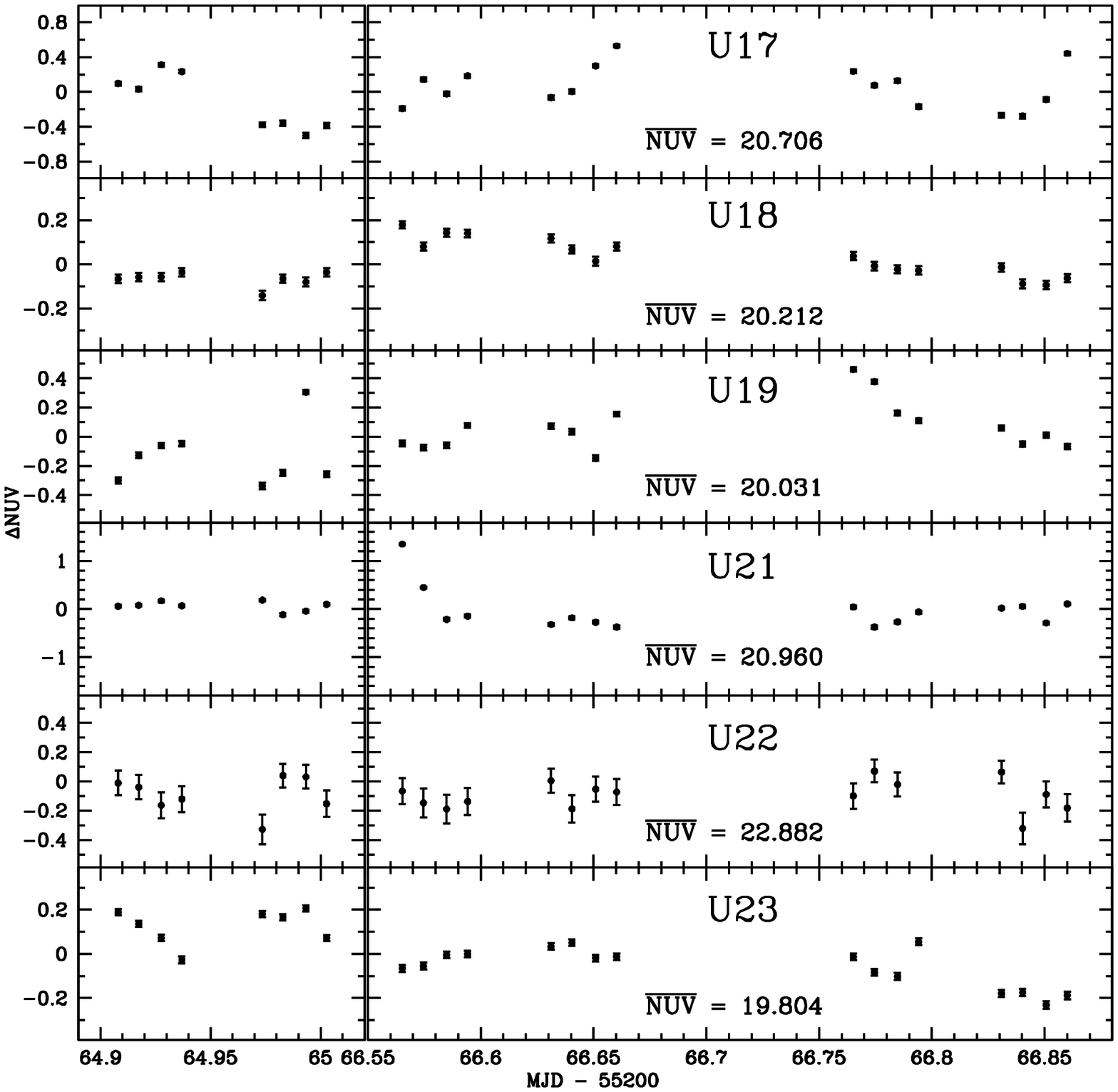}
  \caption{Light curves of the five $NUV$ counterparts to the CVs U17,
    U19, U21, U22 and U23, and the MSP U18 (second panel from the
    top). The $NUV$ observations were carried out on the 9th (first
    two epochs) and 11th (last four epochs) of March 2010. The
    $NUV$ variation from the mean $NUV$ magnitude is plotted versus
    the modified Julian date (MJD). Their mean (average) $NUV$
    magnitude is indicated. See the text for details.}
  \label{variabillity}
\end{figure*}

CVs are known to show all sorts of variable behavior in their light
curves \citep{Warner2003}. Since only four images are available in the
$FUV$ data set, we used the $NUV$ data, which comprised 24 images, to
construct light curves for the five CVs in our $FUV-NUV$ CMD -- U17,
U19, U21, U22, and U23 -- and for the $NUV$ counterpart to the MSP
U18. The $NUV$ observations were taken on the 9th and 11th of March
2010, and all 24 images were obtained with the same exposure time of
680 seconds. All the individual $NUV$ magnitudes were determined with
{\tt Dolphot}. Fig.~\ref{variabillity} shows the light-curves for all
these six sources. We plot the difference between observed magnitude
and mean magnitude, $\Delta NUV = \overline{NUV} - NUV_{obs}$, against
observing time in MJD. Note that the magnitudes assigned to the source
(and hence plotted in the CMDs in Fig.~\ref{cmdfuv} and
\ref{cmdnuvopt}) is the weighted mean (see also Table~\ref{NUVXray}),
which is slightly different to the mean given in the right panel and
used to obtain $\Delta NUV$.

All light curves show brightness variations of a few tenths of a
magnitude on short timescales of minutes. This is typical of the
aperiodic ``flickering'' seen in CVs.

The light curves of the $NUV$ counterparts to the X-ray sources U17,
U19 and U22 all show variation of up to $\approx 0.5$ mag
(semi-amplitude) on short timescales, indicating the typical
flickering variability characteristic of CVs. Towards the end of the
fourth and the beginning of the fifth observing epoch, U17 and U19
show some brightening. However, the epoch in between our fourth and
fifth epoch is not covered by the $NUV$ observations, so we cannot
tell whether the sources were indeed brighter in between the fourth
and fifth epoch or not.  \cite{Shara2005} suggested that the CVs U17
(CV3 in Shara et al.~2005, Grindlay et al.~2001b) and U19 (CV2)
underwent DN eruptions. U17 (CV3) was seen to brighten in optical
WFPC2 data, but not in the STIS $FUV$ data available at that
time. However U19 (CV2) was seen to brighten in the $FUV$ F25SRF2
data, compared to the $FUV$ F25QTZ data. We do not observe any DN
outbursts in our $NUV$ data.

U21 also shows small amplitude variations on short
timescales. \cite{Cohn2010} reported that U21 had the largest
brightening in their H$\alpha$ variability study. The $NUV$
counterpart is brighter by about one magnitude at the beginning of the
third observing epoch, but no further trend is apparent.

U18 and U23 also show somewhat smaller variations of $\approx 0.2$ mag
(semi-amplitude). U23 was classified as a CV, and the light-curve shows
the typical flickering behaviour. U18 is a likely MSP, and its
light-curve shows flickering behaviour, but other than that we do not
see any periodic variability.

\section{Radial Distributions of the Stellar Populations}
\label{radial}

\subsection{The Cluster Center}

In order to study the radial distribution of the stellar populations
that show up in our $FUV-NUV$ and $NUV-V$ CMDs, we used the cluster
centre suggested by \cite{Cohn2010}, at \linebreak RA$= 17^{\rm{h}}
40^{\rm{m}} 42.17^{\rm{s}}$, DEC$= -53^\circ 40\arcmin
28.6\arcsec$. We transformed these coordinates to our $NUV$ reference
frame, using the same transformation that was used to bring the X-ray
coordinates to the $NUV$ reference frame (see section \ref{matching}).

\subsection{Radial Distribution}

Unfortunately, the $FUV$ observations were not centred on the cluster
centre, see Fig.~\ref{FUV_NUV_master_images}. Only a circular area
with a $5.5\arcsec$ radius is fully covered by the $FUV$ data, larger
annuli are only partially covered. However, the $NUV$ and optical data
fully cover a larger circle of $19.8\arcsec$ radius. Therefore we used
the $NUV-V$ CMD in Fig.~\ref{cmdnuvopt} to select the different
stellar populations, i.e. MS stars, BSs, and WDs. We also include
sources that are missing from the $NUV-V$ CMD but are detected in the
$FUV$ data. These include one BS which is above the MS turnoff in the
$FUV-NUV$ CMD but within the MS in the optical CMD, and seven WDs that
can be seen in the $FUV-NUV$ CMD but have no optical counterpart.

HB stars are located within the inner $19.8\arcsec$, but only one has
a $NUV$ (and $FUV$) counterpart (the brightest $FUV$ source, saturated
in the $NUV$, see Sect.~\ref{bs_analysis}). Thus, we selected the HB
stars based on the optical CMD in Fig.~\ref{cmdoptfull}. 

For the radial distribution of the X-ray sources, we used the
catalogue (and coordinates) given in \cite{Cohn2010}. A subset of the
X-ray sources are classified as interacting X-ray binaries in
\cite{Cohn2010}. These include mostly CVs, but also one qLMXB and one
MSP (source U18 discussed above) are located within the annulus
covered by our data. The $FUV$ field of view contains 18 X-ray
sources, but only 14 have a $FUV$ and/or $NUV$ counterpart, five
of those are classified as CVs. U25 is also classified as a CV but
only has a $FUV$ and no $NUV$ counterpart. U18 is a MSP for which we
detected the $NUV$ counterpart, but no $FUV$ match.

The number of sources found in the full $FUV$ field of view, the inner
$5.5\arcsec$ and the inner $19.8\arcsec$ are given in
Table~\ref{num}. Note that the stellar populations and the
corresponding numbers for the full $FUV$ field of view were selected
based on the $FUV-NUV$ CMD, but the number of sources within the inner
$5.5\arcsec$ and $19.8\arcsec$ were selected based on the $NUV-V$ CMD,
and the optical CMD for the HB stars, as outlined above. The number of
all X-ray sources and of the CVs within the inner $5.5\arcsec$ and
$19.8\arcsec$was selected from the catalogue and coordinates presented
in \cite{Cohn2010}, irrespective of them having a counterpart in any
of the other wavebands or not. The CV sample is selected from the
X-ray catalogue and also includes the MSP and the qLMXB, see above.

\begin{table}
\centering
\caption{Number of sources associated with each stellar population. 
  Col. 2 lists the number of sources within the full $FUV$ field of 
  view and based on the $FUV-NUV$ CMD only. Col. 3 lists the numbers within 
  a $5.5\arcsec$ circular area centred on the cluster centre and fully 
  covered by all wavebands, and Col. 4 denotes the number of sources 
  within the inner $19.8\arcsec$ covered by the $NUV$  and optical data. 
  For the latter two columns, the selection is either based 
  on the $NUV-V$ and/or the optical CMD, see the text for details. For the 
  Xray and CV source numbers, we used the coordinates given in Cohn et 
  al.~(2010), but note that not all X-ray sources have a $FUV$, $NUV$ or 
  optical counterpart.}  
\begin{tabular}{cccc}
  \hline
  population & full $FUV$ & $5.5\arcsec$ & $19.8\arcsec$ \\\hline
  MS         & 155 & 83 & 588 \\ 
  BS         & 10  & 8  & 14 \\ 
  HB         & 1   & 0  & 6  \\ 
  X-ray      & 14  & 4  & 25 \\ 
  CV         & 6   & 3  & 7  \\
  WDs        & 16  & 5  & 19 \\ \hline
\end{tabular}
\label{num}
\end{table}

\begin{table}
\centering
\caption{Probabilities in \% returned by the KS test for the comparison of the 
  various stellar populations located in the inner $5.5\arcsec$ circular 
  area (col. 2) and the inner $19.8\arcsec$ (col. 3).}
\begin{tabular}{cccc}
  \hline
  population & $5.5\arcsec$ & $19.8\arcsec$ \\\hline
  MS - BS    & 10.3 & 0.2  \\ 
  MS - WD    & 91.4 & 8    \\ 
  MS - CV    & 50.0 & 0.9  \\ 
  MS - HB    & --   & 40.1 \\ 
  MS - X-ray & 34.7 & 30.1 \\ 
  BS - WD    & 36.6 & 12.2 \\
  BS - CV    & 38.5 & 49.2 \\ 
  BS - HB    & --   & 0.4  \\ 
  BS - X-ray & 74.0 & 1.9  \\ 
  WD - CV    & 94.7 & 39.9 \\ 
  WD - HB    & --   & 6.3  \\ 
  WD - X-ray & 99.4 & 68.3 \\ 
  CV - HB    & --   & 0.8  \\ 
  CV - X-ray & 96.1 & 11.1 \\ 
  HB - X-ray & --   & 22.7 \\\hline 
\end{tabular}
\label{kstest}
\end{table}

\begin{figure}
  \centering
  \includegraphics[width=0.5\textwidth]{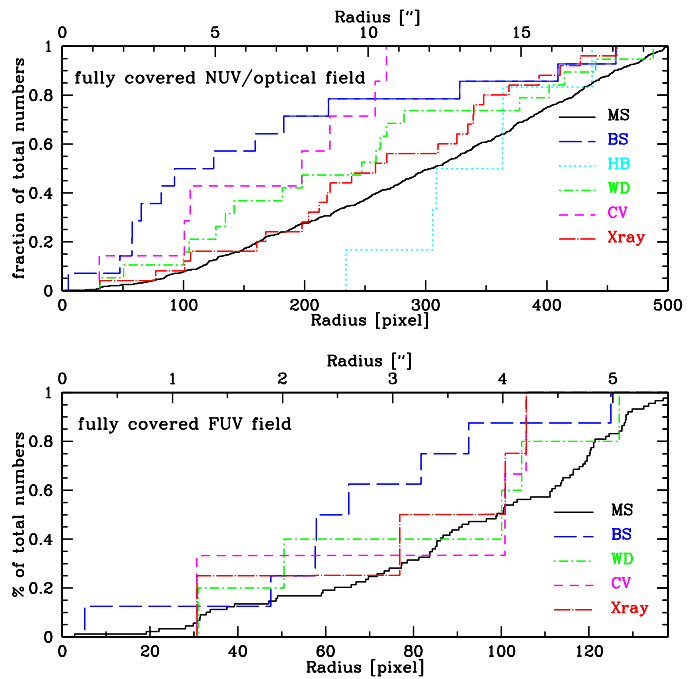}
  \caption{Top panel: Radial distribution of the various stellar
    populations within the inner $19.8\arcsec$ that is fully covered by
    the $NUV$ and optical data but not the $FUV$ data. Bottom panel:
    Radial distribution within the inner $5.5\arcsec$ circular area
    that is also fully covered by the $FUV$ master image.}
  \label{cumulative}
\end{figure}

Fig.~\ref{cumulative} shows the radial distribution of all the stellar
populations discussed above. BSs are the most centrally concentrated
population in both circular areas, while the HB stars are the least
centrally concentrated populations. In fact, HB stars are absent in
the inner circle with $5.5\arcsec$ radius. The CVs (six CVs and the
MSP U18) are concentrated towards the cluster
centre. \cite{Pooley2006} suggested that the CV population in GCs must
have a large dynamically formed component. Based on their location in
the core of the GC, these six CVs might have formed via dynamical
processes. On the other hand, their location might also be the result
of mass segregation, i.e. more massive sources, like CVs and BSs,
travel to towards the cluster core.

We applied the Kolmogornov-Smirnov (KS) test to compare all stellar
populations, see Table~\ref{kstest}. The KS test returns the
probability that the maximum difference should be as large as observed
under the hypothesis that both distributions are drawn from the same
parent distribution. Thus, the higher the probability returned, the
more similar are the two distributions. The smaller the probability,
the less likely it is that the two distributions are drawn from the
same parent distribution, i.e. the more different they are.

Interestingly, X-ray sources are less concentrated than WDs within the
$19.8\arcsec$ circular area. The KS test returns a probability of 68\%
that both the WD and X-ray distributions are drawn from the same
parent distribution. This is not statistically significant. However,
for the inner $5.5\arcsec$, the KS probability is 99.4\%, closer to a
$3\sigma$ significance (99.87\%). If we use the MS sample as the
reference population, then it appears that the X-ray sources are more
centrally concentrated within the inner $5.5\arcsec$ (34.7\%) than the
WDs (91.4\%, more similar to the MS sample), and the CVs are in
between (50\%).

The smallest probability was returned for the larger $19.8\arcsec$
area for the MS vs. BSs sample (0.2\%), BSs vs. HB stars (0.4\%), CVs
vs. HB stars (0.8\%), and MS vs. CVs (0.9\%), indicating the
difference in radial distribution (and hence in mass) between these
populations.

\section{Summary}
\label{sum}

We have carried out a $FUV$ and $NUV$ photometric study on a field
centered on the GC NGC\,6397. 192 $FUV$ sources were detected, of
which 182 have a $NUV$ counterpart. The $FUV-NUV$ CMD, the first
presented for this cluster, shows a sequence of MS stars and 16 WDs
and gap sources. The three brightest $FUV$ sources are saturated in
the $NUV$ so that we cannot clearly classify them as BSs or blue HB
stars based on the $FUV-NUV$ CMD alone. We matched our $FUV-NUV$
catalogue to optical ACS data, which confirmed the classification of
eight BSs. The optical CMD suggests that two of the three brightest
$FUV$ sources are likely bright BSs, resulting in 10 BSs detected in
the $FUV$, but the brightest $FUV$ source is most likely a blue HB
star.

Eighteen X-ray sources are located within our $FUV$ field of view, ten
of those have a $FUV$ counterpart, and 13 have a $NUV$ counterpart,
nine of those have both a $FUV$ and a $NUV$ match. Five (U17, U19,
U21, U22, and U23) of the six CVs in the $FUV$ field of view were
detected in both the $FUV$ and $NUV$. All of these five sources are
located in the WD/gap area in our $FUV-NUV$ CMD, and in the gap
between the WD and MS sequence in the $NUV-V$ CMDs. The sixth CV U25
has only a $NUV$ but no $FUV$ counterpart. We clearly detect the $NUV$
counterpart to the likely MSP U18 \citep{Cohn2010}. A faint ``smudge''
can be seen in the $FUV$ master image at the position of the MSP, but
we did not assign a $FUV$ magnitude. All CVs and the MSP show
short-timescale variability of the order of tens of a magnitude in
their $NUV$ light-curves, i.e. typical flickering behaviour that is
characteristic for CVs. Two of the CVs have previously been reported
to show DN outbursts \citep{Shara2005}, however, none of the CVs
exhibits any large-scale variability indicative of a DN outburst in
our $NUV$ light-curves.

The radial distributions suggest that the BSs and then the CVs are the
most centrally concentrated population in both the inner $5.5\arcsec$
that are fully covered by the $FUV$ data, and a larger $19.8\arcsec$
annulus that is covered by the $NUV$ and optical data. The HB stars,
on the other hand, are the least concentrated population and in fact
absent in the inner $5.5\arcsec$ annulus. The central concentration of
BSs and CVs likely reflects their larger masses, compared to the
smaller masses of the more widely distributed HB stars. The preferred
location of the BSs and CVs towards the cluster centre might be an
effect of mass segregation, i.e. more massive stars travel towards the
cluster centre, and/or of their formation, as BSs and CVs likely form
via dynamical processes in the dense cluster centre.

\section*{Acknowledgements}

A.D. acknowledges support from the People Program (Marie Curie
Actions) of the European Union's Seventh Framework Program
FP7-PEOPLE-2013-IEF under REA grant agreement number 629579. We thank
Jan Pflamm-Altenburg for helpful discussions.






\bsp	
\label{lastpage}
\end{document}